# Analysis of the Shortcut Learning and Clever Hans Effect in CNN based ECG Image Classification

Abhay Kumar Pathak[1,2], Mrityunjay Chaubey[3], Manjari Gupta[1], Deepti Mishra[2]
[1]*Department of Computer Science, Institute of Science, Banaras Hindu University, Varanasi, India.*
[2]*Department of Computer Science (IDI), Norwegian University of Science and Technology (NTNU), Gjøvik, Norway.*
[3]*University of Petroleum and Energy Studies, Dehradun, India.*
**Corresponding Author :** Deepti Mishra (deepti.mishra@ntnu.no ), Manjari Gupta (manjari@bhu.ac.in )

Abhay Kumar Pathak (pathakabhay@bhu.ac.in, abhay.k.pathak@ntnu.no )
Mrityunjay Chaubey (mrityunjay.chaubey@ddn.upes.ac.in )

**Abstract**: Deep learning models for ECG image classification may achieve high accuracy by exploiting non-physiological visual cues instead of ECG waveform morphology. Given the black-box nature of deep learning models, their promise of high predictive performance often remains insufficiently translated into clinical or real-world trust, interpretability, and actionable decision-making. In this study, we examine shortcut learning and Clever Hans effect in a publicly available ECG image dataset using convolutional neural networks. In process we have created six image-derived feature sets (FSs), FS1: raw full ECG images, FS2: cropped waveform-only images, FS3: waveform-masked metadata images, FS4: red-arrow artifact images for the myocardial infarction class, FS5: contrast-enhanced images for the abnormal heartbeat class and FS6: Gaussian-blurred images for the normal class. These controlled representations were used to test whether classification performance persists when waveform information is removed or when artificial class-specific artifacts are introduced. Shortcut retention score, prediction consistency and confidence divergence across Feature-Set Representations have been calculated to assess the transparency about the learning pattern. Along with factual results, average Integrated Gradients and occlusion sensitivity test results are presented to inspect whether model attribution focused on ECG-relevant waveform regions or on non-clinical artifacts. Performance changes across feature sets and attribution patterns were used to identify potential Clever Hans behavior. This study evaluates whether ECG image classifiers learn clinically meaningful morphology or shortcut cues introduced by report layout, metadata, contrast, blur, or artificial markers.

**Keywords**: ECG image classification; Clever Hans effect; shortcut learning; convolutional neural network; explainable AI; Integrated Gradients; medical image analysis.

## Introduction

Last two decade have seen an explosion in the use of AI based system in different aspect of problem solving such as business, healthcare, agriculture and finances [1]. Specially it has been utilized mostly in the medical image analysis [2, 3]. From MRIs to histopathological slides, there are wide spectrum of AI applications have been built, and some have been deployed for gamut of clinical tasks and decision making. In many situations state-of-the-art models are beginning to outperform human experts, with higher volume of data with more computation power, learning based automation and decision-making process predicted to fundamentally change the way clinical medicine operates [4, 5]. However, proceeding with cautiousness, as models are becoming complex and sophisticated, they become increasing inly opaque being difficult to interpret with actual attributes. Development of decision models without having domain specific knowledge and irrelevant learning parameters can lead to enormous results and false conclusions. This phenomenon further leads to shortcut leaning and Clever Hans effect which translates when models start to learn shortcut cues and unintended features in decision making [6, 7]. As developers of models with the potential to significantly affect people's lives, researchers and practitioners must remain vigilant about these issues. The goal should not be limited to developing models with higher predictive performance; rather, it should also involve understanding the basis of model performance, identifying potential limitations, and evaluating the broader consequences of deploying such models in real-world settings [8, 9]. This effect is illustrated several commonly used medical and non-medical problems [10–12]. In a well-known demonstration of Clever Hans-type behavior, Lapuschkin et al. showed that a classifier trained on the PASCAL VOC dataset achieved strong performance across categories such as person, train, car, and horse, but explanation heatmaps revealed that one model was not primarily using the horse itself for prediction. Instead, the model focused on a source tag located in the image corner, which appeared disproportionately in horse images. When the tag was removed, the model's horse prediction weakened; when the same tag was inserted into a car image, the prediction shifted toward horse. This example illustrates that high test accuracy can arise from spurious dataset correlations rather than valid object-level learning. Similar shortcut behavior can occur in medical image models when predictions are driven by non-clinical cues such as text, layout, annotations, or image artifacts rather than disease-relevant morphology [13]. Zhu et al. showed that CNNs can still classify images using background context alone, even when the foreground object is removed. Their background-only model achieved reasonable performance, demonstrating that models may rely on contextual or non-target cues rather than the intended object features [14]. Beery et al. showed that image classifiers can perform well when test images come from familiar contexts but generalize poorly when the same object appears in a novel environment. For example, animal recognition models trained on common camera-trap locations performed strongly within known settings but failed under unseen environmental conditions, showing that models may learn context-specific cues rather than robust object features [15]. Multiple studies have demonstrated that neural networks may use spurious cues in the dataset to achieve apparently strong classification results [9, 16, 17].

In a medical example, Perlich et al. reported information leakage in the KDD Cup breast cancer identification challenge, where patient identifiers carried strong predictive information about tumor malignancy. This showed that models could benefit from non-clinical identifiers rather than relying only on disease-relevant features, leading to inflated performance estimates [18]. Raimondi et al. showed that cancer driver-variant predictors could exploit dataset construction bias by learning to recognize driver genes rather than the functional effects of individual variants. When this shortcut was reduced using a less biased benchmark, model performance dropped by 8–28 AUC percentage points, indicating that the earlier high performance partly reflected shortcut learning rather than true variant-level prediction [19]. DeGrave et al. showed that COVID-19 chest radiograph classifiers often learned image-source-specific shortcuts rather than COVID-relevant pathology. Their models performed worse on external test sets from different sources, and saliency maps showed attention to non-pathological cues such as laterality markers introduced during radiograph acquisition. This demonstrated that apparently strong performance could reflect shortcut learning rather than clinically meaningful disease detection [20]. Maguolo and Nanni showed that COVID-19 X-ray classifiers could achieve similar performance even when large zero-intensity regions were placed over the lungs, forcing the model to learn mainly from the outer parts of the image. This indicated that some models were exploiting dataset-specific biases rather than learning COVID-19-related lung features [21]. Several ECG image-classification studies have reported high model performance using deep learning architectures; however, these studies generally evaluated performance using conventional metrics and did not explicitly test whether the models relied on shortcut cues or Clever Hans-type behavior. For example, Sadad et al. used 12-lead ECG images and reported efficient CNN-based classification across four cardiac categories, while Ao et al. demonstrated the feasibility of CNN-based diagnosis directly from raw 12-lead ECG images. More recent ECG image studies have also focused on classification performance and robustness, but systematic evaluation of whether predictions are driven by ECG waveform morphology or non-physiological visual cues remains limited [22–24]. Czerwinski et al. used scanned ECG images for arrhythmia detection and AF recurrence prediction, showing that ECG scans can be treated as image-based clinical inputs for deep learning. Their study also emphasized interpretability and cross-dataset validation, reporting that performance dropped when models trained on one dataset were tested on another, indicating domain-shift challenges in ECG scan analysis. This supports the relevance of evaluating not only classification accuracy but also model behavior and explanation patterns in ECG image-based AI systems [25].
In this study, we demonstrated Clever Hans and shortcut learning affecting classification performance in publicly available ECG image dataset. The dataset consists of four classes, originally developed made available and used by Khan and Hussain [26] (https://data.mendeley.com/datasets/gwbz3fsgp8/2). To assess representation-dependent performance shifts, we created six ECG feature-set representations. The first three feature sets were designed to separate the main information sources present in the ECG report image: Feature Set 1 represented the raw full ECG image, Feature Set 2 represented the waveform-only ECG image, and Feature Set 3 represented the metadata-only or waveform-masked ECG image. The last three feature sets were designed as artifact-induced conditions in specific classes: Feature Set 4 represented the red-arrow condition, Feature Set 5 represented the contrast-enhanced condition, and Feature Set 6 represented the Gaussian-blurred condition. These feature sets were used to examine whether CNN performance remained driven by ECG waveform morphology or shifted under non-waveform and artifact-induced visual conditions.

**The contributions of this study are as follows:**

- A controlled feature-set framework was developed to assess shortcut learning and Clever Hans-type behavior in CNN-based ECG image classification.
- Six ECG representations were evaluated to separate the effects of waveform morphology, metadata/report-context information, and artifact-induced visual cues.
- Shortcut Retention Score, prediction consistency, and confidence divergence were used to quantify representation-level changes in model behavior.
- Integrated Gradients and occlusion sensitivity were used to visually and quantitatively examine whether CNN decisions were driven by ECG morphology or non-physiological shortcut cues.

## Mathematical Preliminary:

This section includes the mathematical background of CNN model and explainable AI techniques that have been utilized in this study.

**CNN:** Convolution-based deep leaning model leans spatial dependency by applying convolutional operation along with height and weight dimensions where representation at each spatial locations are computed using fixed location neighbourhood using shared kernels.
Given a temporally ordered sequence $I = \mathbb{R}^{H \times W \times C}$ where $H, W$ and C are height, width and the channel of the image. convolution-based models construct spatial representations through two-dimensional convolutions. A two dimensional convolution is defined as

$$h_{u,v} = \sum_{p=0}^{K_h-1} \sum_{q=0}^{K_w-1} W_{p,q} I_{u+p,v+p}$$

Where $K_h$ and $K_w$ represents kernel height and kernel width respectively and $W_{p,q}$ are the learnable convolutional parameters. Where $I_{u+p,v+p}$ represents the input pixel at spatial position $(u+p, v+p)$ and $h_{u,v}$ denotes the hidden representation at spatial location $u, v$.

For a deep CNN architecture composed of multiple convolutional layers, the representation at layer $l$ is given by:

$$h_{u,v}^{l} = \sigma(\sum_{p=0}^{K_h-1} \sum_{q=0}^{K_w-1} W_{p,q}^{l}\, h_{u+p,v+p}^{l-1} + b^{l})$$

Where $h_{u,v}^{l}$ represents the spatial representation at location $(u, v)$ in layer $l$, $h_{u+p,v+p}^{l-1}$ represents the local neighbourhood from previous layer $l-1$, $W_{p,q}^{l}$ are the learnable convolution parameter, $b^{l}$ is the bias term and $\sigma(*)$ is a non-linear activation function.

The output at each time step is computed as $\hat{y} = g(h)$, $g(*)$ denotes the classification function applied after convolution feature extraction.

where and the model parameters are optimized by minimizing the sequence-level loss:

$$\mathcal{L} = l(y, \hat{y})$$

using standard backpropagation. Where $\mathcal{L}$ denotes the training loss and function $l(y, \hat{y})$ is the loss function comparing the true output $y$ and the predicted output $\hat{y}$.

**Integrated Gradients (IG) :** It is a gradient-based explainable artificial intelligence technique that attributes the prediction of a deep sequential model to its input features by integrating gradients along a path from a baseline input to the actual input. In the context of ECG image classification, IG can be used to identify the spatial regions of an ECG image that most influence the prediction of a convolutional neural network.

Consider a trained CNN-based classification model:

$$f_c\colon \mathbb{R}^{H \times W \times C} \rightarrow \mathbb{R}$$

Where H, $W$ and C are height width and number of image channels respectively and $f_c(I)$ represents the model output score for class c. And an image input be:

$$I = \mathbb{R}^{H \times W \times C}$$

Let $I'$ be the baseline image.

$$\mathrm{I}^{(\alpha)} = I' + \alpha(\mathrm{I} - I'), \in [0,1]$$

The attribution pixel location (u, v) and channel $k$ I is computed as :

The attribution feature $j$ at time step $t$ is

$$IG_{u,v,k}(I) = (I_{u,v,k} - I'_{u,v,k}) \int_0^1 \frac{\partial f_c(I^{\alpha})}{\partial I_{u,v,k}} d\alpha$$

Using $M$ discrete aprroximation steps, this becomes

$$IG_{u,v,k}(I) = (I_{u,v,k} - I'_{u,v,k}) \frac{1}{M} \sum_{m=1}^{M} \frac{\partial f_c(I' + \frac{m}{M}(I - I'))}{\partial I_{u,v,k}}$$

The attribution map

$$IG(I) \in \mathbb{R}^{H \times W \times \mathrm{C}}$$

Which indicates contribution of each pixel or spatial region of the ECG image towards CNN prediction.

$$IG_{u,v}(I) = \frac{1}{C} \sum_{k=1}^{C} \left| IG_{u,v,k}(I) \right|$$

Thus, Integrated Gradients provides fine-grained spatial interpretability for CNN-based ECG image classification by identifying which pixels, waveform regions, or lead-specific areas most influence the predicted class. For global explainability, IG attributions can be aggregated across multiple ECG images belonging to the same class. Let:

$$D_c = \{I^1, I^2, \dots, I^{\mathrm{N}_c}\}$$

Be the set of images belonging to the class $c$. The class wise global attribution is computed as:

$$\overline{IG_{u,v,k}^{(c)}} = \frac{1}{N_c} \sum_{i=1}^{N_c} IG_{u,v,k}(I^i).$$

To summarize importance across time and features, aggregated measures can be defined as:

$$\overline{IG_{u,v}^{(c)}} = \frac{1}{C} \sum_{k=1}^{C} \left| IG_{u,v,k}^{(c)} \right|$$

These aggregated attributions reveal class-specific spatial relevance patterns and dominant ECG image regions, thereby extending Integrated Gradients from local image-level interpretation to global, class-wise explanation of CNN-based ECG image classification behavior.

**Occlusion Sensitivity (OS) Test:** Occlusion Sensitivity Test is a perturbation-based explainable AI method that evaluates the importance of image regions by masking small parts of the input image and observing the change in the CNN prediction.
Consider the trained classification model:

$$f_c \colon \mathbb{R}^{H \times W \times \mathrm{C}} \to \mathbb{R}$$

And the input ECG image

$$I = \mathbb{R}^{H \times W \times \mathrm{C}}$$

Where H, $W$ and C are height width and number of image channels respectively. A small region $R_{u,v}$ of the image is occluded using a mask value such as zero, mean intensity, or a blurred patch. The occluded image is denoted as: $I_{OCC}^{u,v}$.

The occlusion sensitivity score for class $c$ is computed as:

$$OS_{u,v}^{c} = f_c(I) - f_c(I_{OCC}^{u,v})$$

where $f_c(I)$ is the original class score and $f_c(I_{OCC}^{u,v})$ is the class of occlusion. A higher value of $OS_{u,v}^{c}$ indicates that the masked region has a stronger contribution to the CNN prediction.

By sliding the occlusion mask across the ECG image, an occlusion sensitivity heatmap is obtained:

$$OS^{c}(I) \in \mathbb{R}^{H\times W}$$

For class-wise global explanation, occlusion maps are averaged over multiple ECG images from the same class. Let:

$$D_c = \{I^1, I^2, \dots, I^{N_c}\}$$

be the set of ECG images belonging to class $c$. The average class-wise occlusion sensitivity map is computed as:

$$\overline{OS_{u,v}^{(c)}} = \frac{1}{C}\sum_{k=1}^{C} \left|OS_{u,v,k}^{(c)}(I')\right|$$

where $N_c$ is the number of ECG images in class $c$. This averaged heatmap highlights the common ECG image regions, waveform areas, or lead-specific patterns that consistently influence the CNN prediction for that class.

**Prediction Consistency and Confidence Divergence Analysis Across Feature-Set Representations:** To evaluate whether the CNN classification model produced stable predictions across different ECG image feature-set representations, we performed a representation-level prediction stability analysis. This analysis compares model outputs obtained from multiple feature sets $Feature\ set_1$–$Feature\ Set_6$, where each feature set represents a different form of ECG image information, such as raw ECG, waveform-only ECG, metadata-masked ECG, artifact-modified ECG, contrast-enhanced ECG, and Gaussian-blurred ECG.

Two complementary measures were used: Prediction Consistency Score (PCS) and Jensen–Shannon Divergence (JSD). PCS measures the agreement between predicted class labels from two feature-set representations, whereas JSD measures the divergence between their predicted class-probability distributions.

For two Feature Sets $F_a$ and $F_b$, the Prediction Consistency Score is defined as:

$$PCS(F_a, F_b) = \frac{1}{N}\sum_{i=1}^{N} \mathbb{I}(\hat{y}_i^{F_a} = \hat{y}_i^{F_b})$$

where $N$ is the number of test samples, $\hat{y}_i^{F_a}$ and $\hat{y}_i^{F_b}$ are the predicted class labels for the $i^{th}$ sample under feature sets $F_a$ and $F_b$, and $\mathbb{I}(\cdot)$ is the indicator function. A higher PCS value indicates stronger agreement in predicted class labels.

To quantify confidence-level changes, Jensen–Shannon Divergence was computed between the predicted probability distributions:

$$JSD(P^{F_a}, P^{F_b}) = \frac{1}{2}KL(P^{F_a} \parallel M) + \frac{1}{2}KL(P^{F_b} \parallel M)$$

Where

$$M = \frac{1}{2}(P^{F_a} + P^{F_b})$$

Here, $P^{F_a}$ and $P^{F_b}$ denote the predicted class-probability distributions from two feature-set representations, and $KL(\cdot)$ denotes Kullback–Leibler divergence. Lower JSD values indicate similar confidence distributions, whereas higher values indicate greater confidence divergence.

## Materials and methods

This section includes dataset description, preprocessing and feature set creation, model utilized in the study. The flow diagram of the proposed methodology is provided in **Fig. 1**.

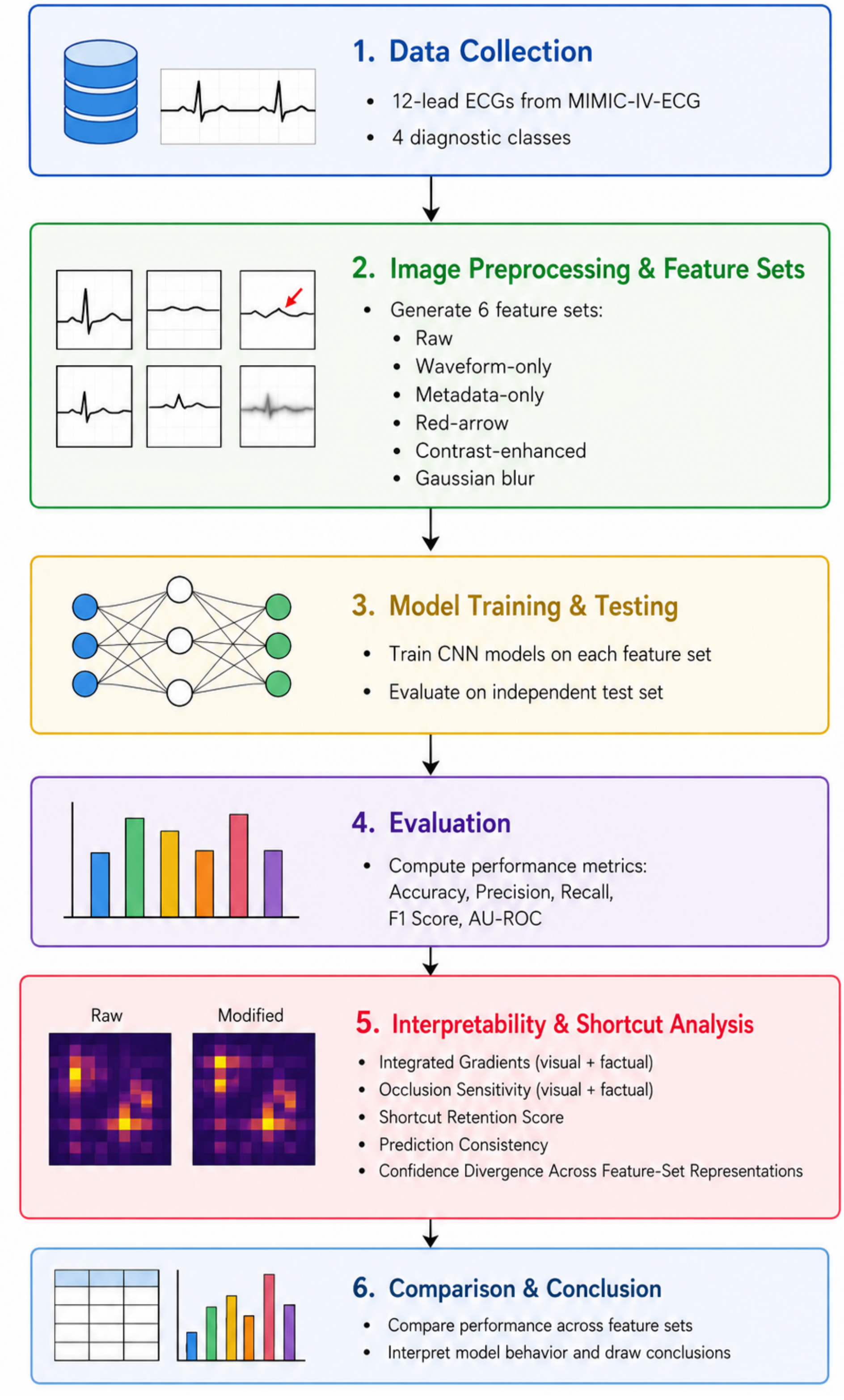

***Fig. 1***: *Workflow of the proposed methodology.*

**Dataset**: This study uses the ECG Images Dataset of Cardiac Patients, available through Mendeley Data. The dataset was published as Version 2 on 19 March 2021 and assigned DOI 10.17632/gwbz3fsgp8.2 [26]. The dataset page describes it as an ECG image dataset of cardiac patients created under the auspices of Ch. Pervaiz Elahi Institute of Cardiology, Multan, Pakistan, with the aim of supporting cardiovascular disease research. The dataset is organized in class-wise folders. Each image corresponds to a scanned or exported ECG report containing ECG waveforms and report-level information. The visual structure of the ECG reports includes waveform traces, lead labels, ECG grid background, calibration pulses, patient/report metadata, acquisition settings, and timestamp information. For the present study, the class-wise folder organization is used to assign labels automatically. Images are resized to a fixed input resolution of 224 × 224 pixels and normalized to the range [0,1]. The dataset is split into training and testing subsets using stratified sampling to preserve the class distribution.

**Feature Set Creation:** Six sets of features based on ECG image representation were extracted from the same original dataset of ECG images as presented in **Table 1**. The same images, the same class labels and the same train-test split were present in each feature set. Only the image content has been modified. The aim was to assess whether the CNN learned clinically meaningful ECG waveform morphology or relied on shortcut cues like metadata, artificial markers, contrast differences, or image-quality artefacts. While creating different feature sets, we have used class number to represent four different classes; Class 1: Myocardial Infarction; Class 2: History of Myocardial Infarction; Class 3: Abnormal Heartbeat; Class 4: Normal Patient; to interpret the results clearly.

| Feature Set | Image Representation / Modification | Class Affected by Modification | Type of Bias Tested | Explanation |
| --- | --- | --- | --- | --- |
| **FS1** | **Raw full ECG image** | **All four classes** | Baseline / natural dataset bias | Uses the complete ECG report to assess baseline performance and possible natural shortcuts. |
| **FS2** | **Cropped waveform-only ECG image** | **All four classes** | Morphology-focused control | Retains only ECG waveform regions to test clinically relevant morphology learning. |
| **FS3** | **Waveform-masked / metadata-only ECG image** | **All four classes** | Metadata / report-context bias | Masks waveform regions to test whether report context alone supports classification. |
| **FS4** | **Raw ECG image with red arrow in upper-left region** | **Myocardial Infarction patient class** | Explicit artificial marker bias | Tests whether the CNN learns a visible class-specific marker instead of ECG morphology. |
| **FS5** | **Contrast-enhanced ECG image** | **Abnormal Heartbeat patient class** | Pixel-intensity / contrast bias | Tests whether class-specific contrast changes act as shortcut cues. |
| **FS6** | **Gaussian-blurred ECG image** | **Normal patient class** | Image-quality / blur bias | Tests whether blur or image sharpness differences influence model prediction. |

***Table 1****: Morphology-Based and Class-Specific Artifact-Induced ECG Image Feature Sets for Shortcut Learning Analysis.*

**General Image Standardization:** All ECG images were loaded in RGB format and resized to the fixed input dimension required by the CNN architecture. Pixel normalization was applied using the same preprocessing method across all feature sets. No feature-set-specific normalization was performed. The original class label of each image was retained after transformation. **Feature Set 1 (Fig.2(a))**: The original full ECG report image without any modification. No cropping, masking, marker insertion, contrast adjustment or blurring was applied. This feature set preserved waveform traces, grid background, lead labels, report text, margins and meta data regions. **Feature Set 2 (Fig. 2(b))**: It is obtained by cropping the ECG waveform region from the complete ECG report image. The 12-lead waveform area was kept in a fixed rectangular region of interest. Excluded were header information, footer information, patient/report metadata, printed diagnostic text, and peripheral report regions. The cropped waveform image was resized to the size of the input to the CNN. **Feature Set 3 (Fig. 2(c))**: It was acquired masking the area of the ECG waveform, while the rest of the report was left unchanged. The same part of the waveform as that used for F2 was selected and replaced with a constant background color matching the background of the ECG sheet. This eliminated ECG morphology but kept report layout, margins, lead labels, printed text, metadata fields and other non-waveform visual information.

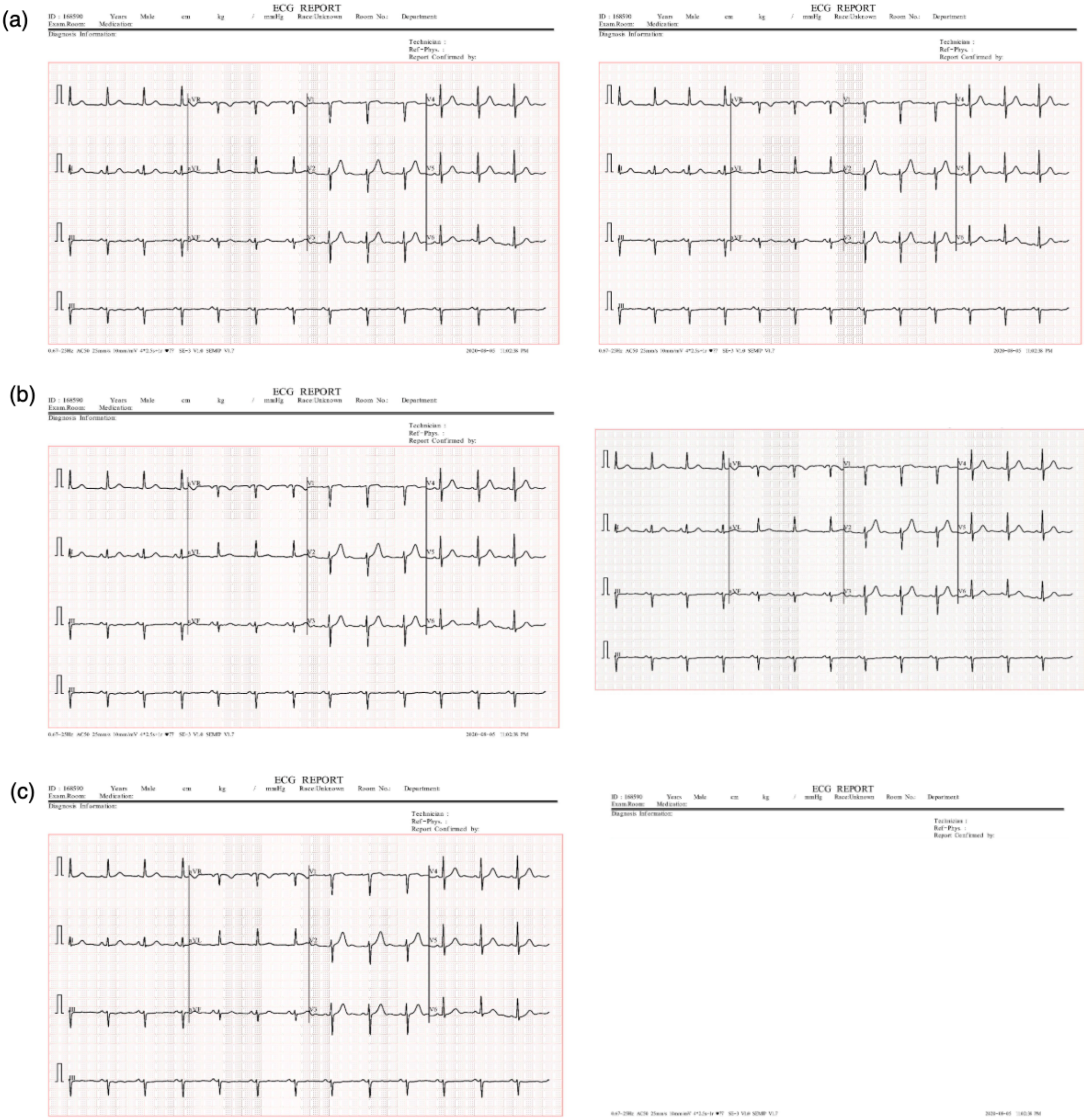

***Fig. 2****: Morphology-Based feature sets; a) Raw Image; b) Waveform only image; c) Metadata only image.*

**Feature Set 4 (Fig. 3 (a))**: The created feature set consists of a synthetic red arrow embedded solely within the myocardial infarction ECG images. The arrow was placed in the same upper-left location in each modified image. The size, color, thickness and orientation of the arrows were held constant. All other classes of ECG images were not altered. The marker was placed outside the main region of the waveform so that the morphology of the ECG was not directly modified. **Feature Set 5 (Fig. 3 (b))**: It was generated by contrast enhancement on abnormal heartbeat ECG images only. All images of abnormal heartbeat were acquired with the same contrast parameters. Normal, myocardial infarction images, and history of myocardial infarction classes were preserved. This induced a controlled class-specific intensity distribution shift. **Feature Set 6 (Fig. 3 (c))**: Generated by Gaussian blur on normal ECG images only. All the normal images were processed with the same Gaussian kernel size and standard deviation. The images of the other classes were not blurred. This resulted in a controllable class-specific image-quality degradation.

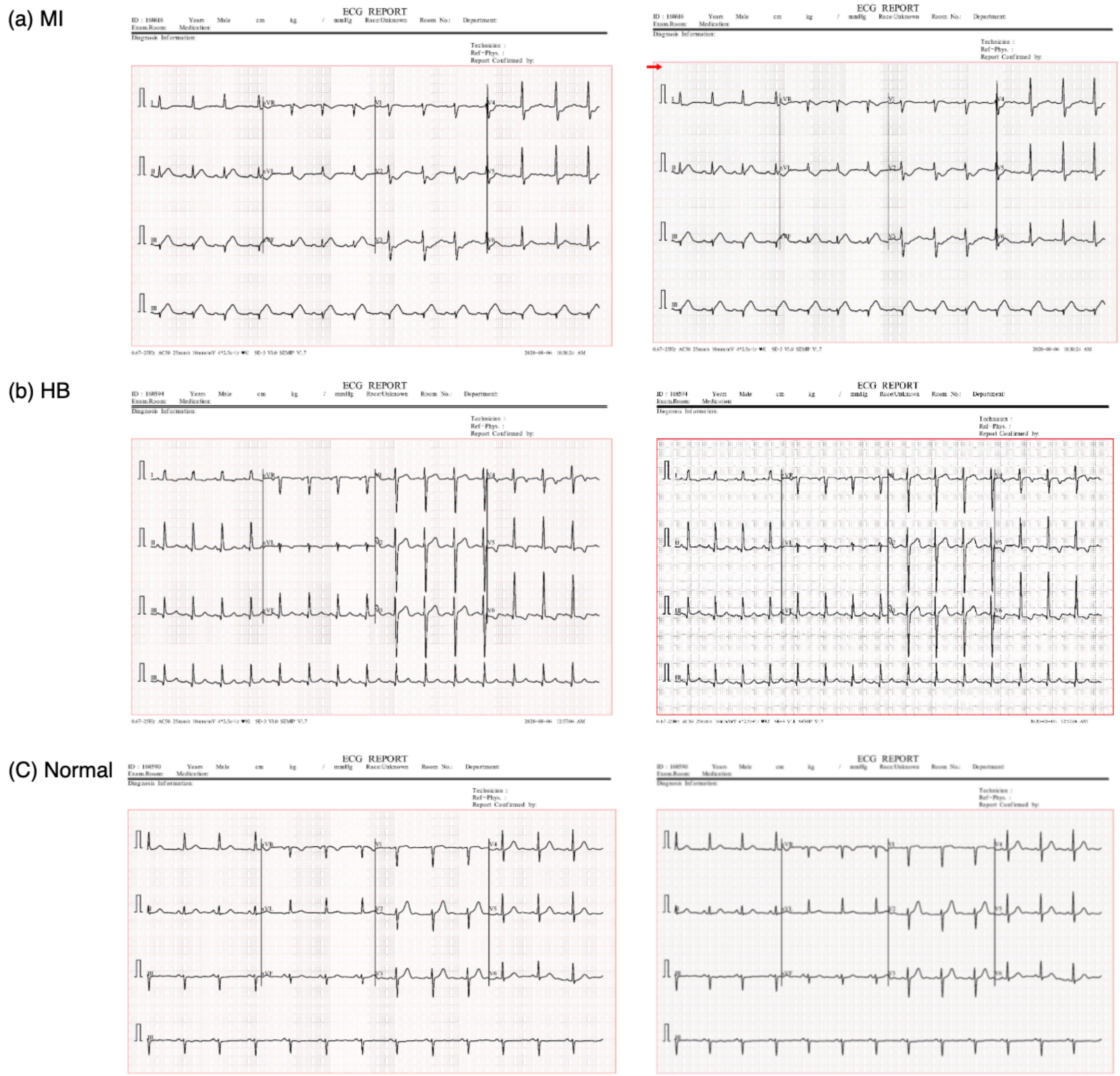

***Fig. 3**: Artefacts induced feature set; a) Red arrow induced in class 1 at left upper side; b) Contrast increment in class 3; c) Gaussian blur induced in class 4.*

**Model Utilized in the Study:** We trained a CNN model (**Fig, 4**) on each ECG image feature set separately, using the same experimental setup. The class labels were assigned automatically according to the class-wise folder structure. All images were loaded in RGB format, resized to 224 × 224 pixels and normalized to [0,1]. The data set was divided into training and test sets using stratified sampling, with 75% of images for training and 25% for testing. For reproducibility we used a fixed random seed of 42. Class imbalance was addressed using balanced class weights based on training labels. The CNN architecture consisted of 5 convolutional blocks with 32, 64, 128, 256 and 512 filters respectively. Each block was composed of a 3 × 3 convolution, ReLU activation, batch normalization and max pooling. The last convolutional block was followed by global average pooling Classification head consisted of 2 dense layers with 256 and 128 neurons with dropout rates of 0.5 and 0.3. The last layer uses SoftMax activation for multi-class classification. The model was trained for 30 epochs using Adam optimizer with a learning rate of 1 x 10 ^{-4} and sparse categorical cross-entropy loss. We measured performance on the independent test set using accuracy, loss, precision, recall, F1-score and confusion matrix analysis. All feature sets were trained and evaluated with the same pipeline to allow for a fair comparison.

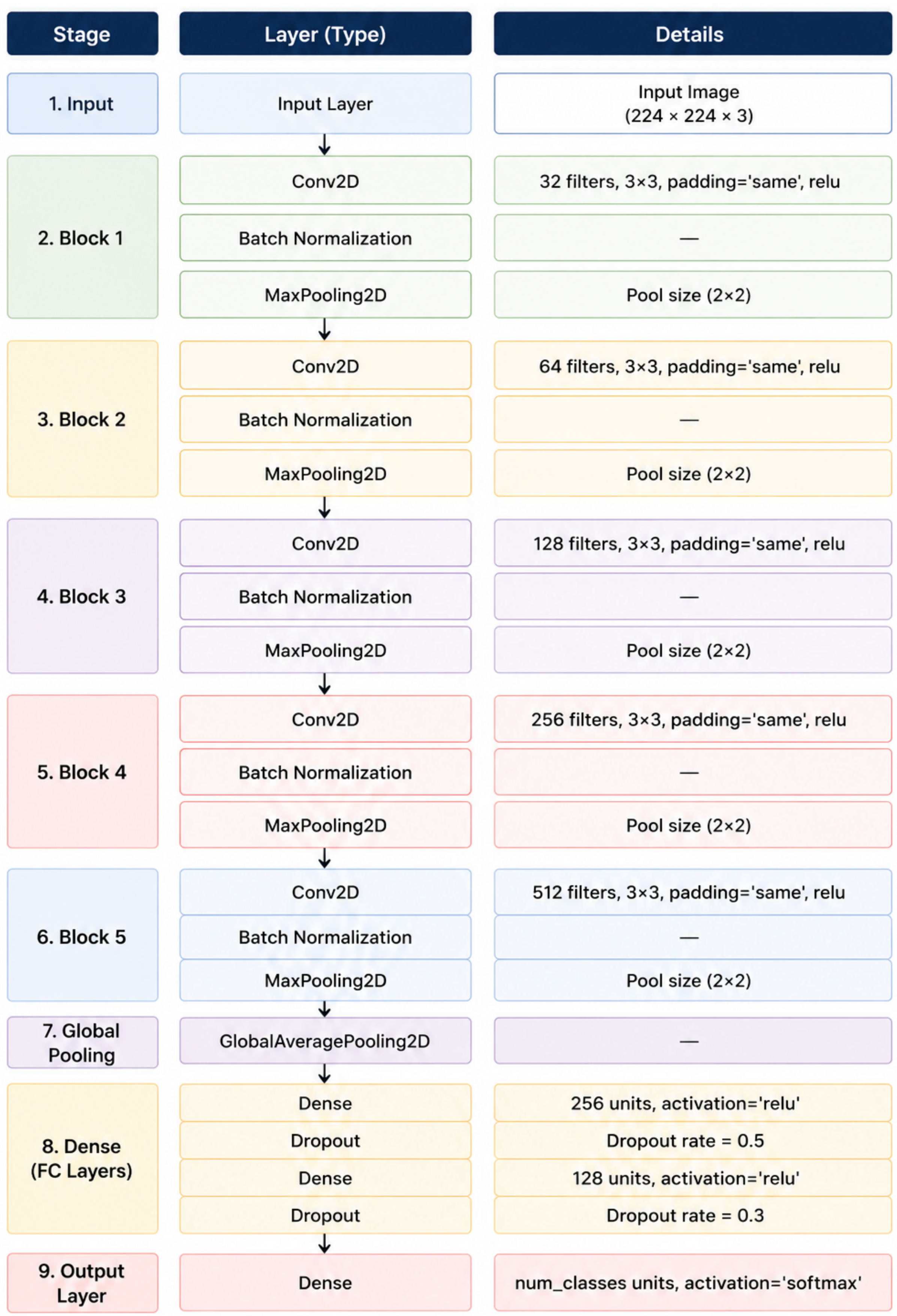

| Stage | Layer (Type) | Details |
|---|---|---|
| 1. Input | Input Layer | Input Image (224 × 224 × 3) |
| 2. Block 1 | Conv2D | 32 filters, 3×3, padding='same', relu |
| | Batch Normalization | — |
| | MaxPooling2D | Pool size (2×2) |
| 3. Block 2 | Conv2D | 64 filters, 3×3, padding='same', relu |
| | Batch Normalization | — |
| | MaxPooling2D | Pool size (2×2) |
| 4. Block 3 | Conv2D | 128 filters, 3×3, padding='same', relu |
| | Batch Normalization | — |
| | MaxPooling2D | Pool size (2×2) |
| 5. Block 4 | Conv2D | 256 filters, 3×3, padding='same', relu |
| | Batch Normalization | — |
| | MaxPooling2D | Pool size (2×2) |
| 6. Block 5 | Conv2D | 512 filters, 3×3, padding='same', relu |
| | Batch Normalization | — |
| | MaxPooling2D | Pool size (2×2) |
| 7. Global Pooling | GlobalAveragePooling2D | — |
| 8. Dense (FC Layers) | Dense | 256 units, activation='relu' |
| | Dropout | Dropout rate = 0.5 |
| | Dense | 128 units, activation='relu' |
| | Dropout | Dropout rate = 0.3 |
| 9. Output Layer | Dense | num_classes units, activation='softmax' |

***Fig. 4**: CNN based model used in the study.*

## Experimental Results and Discussion

In this section we have provided the details about experimental setup and its parameters following with experimental results and discussion.

**Experimental Setup:** CNN is used as a baseline classifier, and the specification of the models has already been discussed in the previous section. Apart from the protocol setup for the feature set creation, there was no preprocessing step involved. For the training and testing of the model on different feature sets, an 80:20 train-test split ratio is applied throughout the experiments. The samples taken for training and testing uses stratified sampling to draw a balanced number of class samples. All the models were implemented and tested on a GPU-based HP Z series workstation having a RAM capacity of 32 GB and 16 GB of GPU memory.

To quantify the extent to which predictive performance was retained in each feature representation compared to the morphology-focused condition, we calculated a Shortcut Retention Score (SRS), with Feature Set 2 as the baseline. Feature Set 2 corresponded to the waveform-only ECG image condition and was thus treated as the morphology-focused baseline.

For each feature set $F_k$, SRS was calculated as:

$$SRS_k = \frac{Acc(F_k)}{Acc(F2)}$$

where $Acc(F_k)$ denotes the testing accuracy of the CNN model trained and evaluated on feature set $F_k$, and $Acc(F2)$ denotes the testing accuracy of the waveform-only feature set. Prediction consistency and confidence divergence were calculated over the entire test set comprising 25% of the overall samples. For each test sample, predictions were obtained over the feature-set representations and used to build two matrices: one measuring prediction agreement over feature sets and the other measuring divergence in confidence-levels. These matrices were then combined into a compact representation that summarized the stability or variability of the CNN predictions over different ECG image representations.

After model training, class-wise attribution maps for each of the feature sets were generated using Integrated Gradients. For each class of ECG, 20 test samples were selected, and IG was calculated with a black baseline with 5 steps of interpolation. The absolute channel-wise attribution values were summed and resized to a $400 \times 400$ matrix, smoothed by Gaussian filtering, normalised to $[0,1]$ and averaged over samples to get a representative IG heatmap for each class. Brighter areas are higher average attribution magnitude for the model prediction.

Following model training on the full test set, visual and quantitative assessment with occlusion sensitivity was performed. Twenty test samples were selected for each class of ECG, and local image regions were occluded with a $24 \times 24$ white patch with a stride of 10. Class-wise sensitivity maps were made using the change in true-class confidence after each occlusion. The resulting maps were resized to $400 \times 400$, smoothed with a Gaussian kernel, normalised and visualised as heatmaps. Quantitative values were extracted including Mean OS, Strong OS, Top 10% concentration, localization and prediction flip rate.

**Baseline Results of CNN Model on Different Feature Set:** In **Table 2,** it shows the training and testing performance of the CNN models over the six ECG feature sets. Among all representations, Feature Set 5 performed the best on testing with accuracy 0.9655, precision 0.9710, recall 0.9565, F1 score 0.9623 and AU-ROC 0.9930. The waveform-only representation, Feature Set 2, also surpassed the raw ECG baseline with a testing accuracy of 0.8060 compared to 0.7629 for Feature Set 1. In contrast, the lowest testing performance was observed in the Feature Set 3 where the waveform information was masked and only the information of the meta data/report-context was available with an accuracy of 0.5431 and F1 score of 0.4948. The testing performance of Feature Sets 4 and 6 was similar to that of the raw ECG baseline, which indicates relatively smaller performance changes under the red-arrow and Gaussian-blur conditions.

| Feature Set | Training/Testing | Accuracy | Precision | Recall | F1 Score | AU-ROC |
|---|---|---|---|---|---|---|
| Feature Set 1 | Training | 0.824713 | 0.878486 | 0.828263 | 0.796080 | 0.999997 |
| | Testing | 0.762931 | 0.827033 | 0.769394 | 0.735066 | 0.983527 |
| Feature Set 2 | Training | 0.908046 | 0.916027 | 0.908571 | 0.894561 | 1.000000 |
| | Testing | 0.806034 | 0.828802 | 0.800170 | 0.778565 | 0.982007 |
| Feature Set 3 | Training | 0.597701 | 0.509278 | 0.656425 | 0.533841 | 0.999189 |
| | Testing | 0.543103 | 0.477888 | 0.593134 | 0.494815 | 0.922612 |
| Feature Set 4 | Training | 0.79454 | 0.867198 | 0.793167 | 0.756241 | 0.999833 |
| | Testing | 0.75431 | 0.812990 | 0.740778 | 0.693925 | 0.983470 |
| Feature Set 5 | Training | 1.000000 | 1.000000 | 1.000000 | 1.000000 | 1.000000 |
| | Testing | 0.965517 | 0.970985 | 0.956496 | 0.962264 | 0.992996 |

| Feature Set 6 | Training | 0.788793 | 0.864938 | 0.790000 | 0.727626 | 1.000000 |
|---|---|---|---|---|---|---|
| | Testing | 0.788793 | 0.864034 | 0.788793 | 0.725546 | 0.991566 |

***Table 2**: Training and testing performance of CNN models across six ECG feature sets.*

These changes are further summarized in the relative testing performance plot presented in **Fig. 5**, using Feature Set 1 as baseline. Feature Set 5 showed the largest positive shift in all four metrics (accuracy, precision, recall an d F1 score) compared to the raw ECG condition, indicating a strong performance increase under the contrast-enh anced condition. Feature Set 3 showed the most negative shift in all major metrics, supporting the conclusion tha t removal of waveform morphology significantly decreased classification performance. Feature Set 2 showed a s light positive shift. Feature Sets 4 and 6 hovered around the baseline, with slight differences only at the metric le vel. The table (**Table 2**) and figure (**Fig. 5**) together suggest that the model performance is highly dependent on t he image representation, with the highest performance gain for contrast enhancement and the largest performanc e reduction for the metadata-only representation.

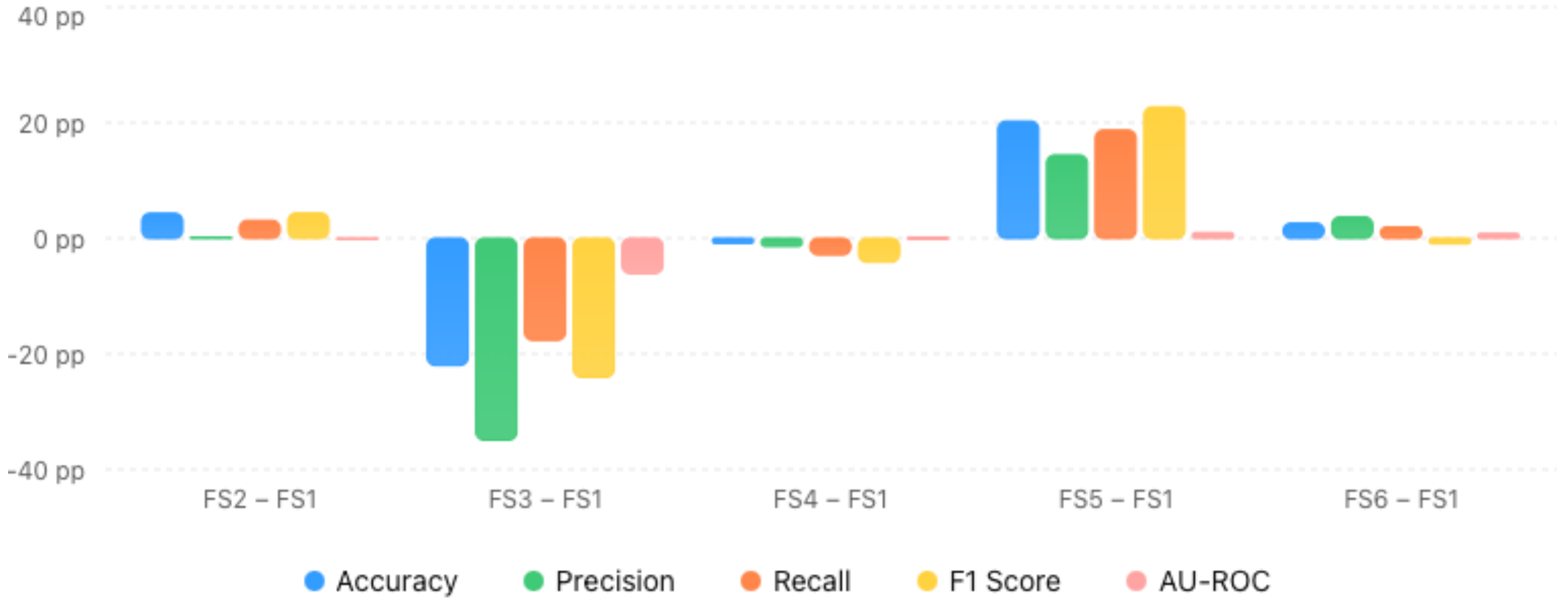

***Fig. 5**: Performance difference across ECG features sets relative to the raw ECG baseline. Values represent percentage-point differences calculated as $\Delta = Feature\ Set_k - Feature\ Set_1$. Positive values indicate higher performance than F1, whereas negative values indicate lower performance than F1.*

**XAI Results:** In this subsection provides the various factual and visual results supporting the explainability analysis of CNN model.

**Shortcut Retention Score:** The shortcut retention score presented in **Table 3**, where all the feature sets are compared with the feature set 2 (waveform only images). Raw full images retained 94.7% of waveform-only accuracy, showcasing that the full ECG images perform poor, leading to presence of shortcut. While Metadata-only images retained 67.4% of baseline performance, indicating that non-waveform report information still contains learnable class-associated signal.

| Feature set | Image representation | Test accuracy | SRS vs F2 |
|---|---|---|---|
| F1 | Raw full ECG image | 0.7629 | 0.947 |
| F2 | Waveform-only ECG image | 0.8060 | 1.000 |
| F3 | Metadata-only / waveform-masked image | 0.5431 | 0.674 |
| F4 | Red-arrow condition | 0.7543 | 0.936 |
| F5 | Contrast-enhanced condition | 0.9655 | 1.198 |
| F6 | Gaussian-blurred condition | 0.7888 | 0.979 |

***Table 3**: Shortcut Retention Score Across ECG Feature Sets Using Waveform-Only Accuracy as the Reference. (Note: SRS = 1.000 means the feature set performs the same as waveform-only ECG images; SRS < 1.000 means the feature set performs worse than waveform-only ECG images; SRS > 1.000 means the feature set performs better than waveform-only ECG images.)*

In artifacts induced features (Feature Set 1, Feature Set 2 and Feature Set 3), shortcut retention score was keeps changing. In feature set 4 and Feature Set 6, it was close to waveform only images suggested remained highly performant under special conditions (class-specific artifacts induced feature sets) while features set 5 exceeded baseline accuracy by 19.8%, indicating that strong performance gain acted as a highly learnable cue.

**Prediction Consistency and Confidence Divergence Across Feature-Set Representations:** The quantitative results of PCS and JSD comparison across various feature sets is presented in Table 4. Using F2 as the waveform-only morphology reference as a baseline we have observed that PCS and JSD changes throughout different feature sets while comparing class and probability of the test across all feature set of the similar test samples. Feature Set 1 showed moderate change in prediction consistency but significant shift in confidence. The highest difference was observed when compared with metadata only images (Feature Set 3) which presented lowest consistency in predicting classes and strongest representation dependent shift. When red arrow artefacts (Feature Set 4) were introduced to specific class, resulted in unstable prediction along with moderate confidence shift. It shows that even class-specific biasness such as 'contrast enhancement' and 'gaussian blur' did not strongly destabilize prediction relative to the waveform only images.

| Reference feature set | Compared feature set | Representation comparison | PCS | JSD |
|---|---|---|---|---|
| F2 | F1 | Waveform-only ECG vs raw full ECG | 0.5759 | 0.3102 |
| F2 | F3 | Waveform-only ECG vs metadata-only / waveform-masked ECG | 0.5131 | 0.4417 |
| F2 | F4 | Waveform-only ECG vs red-arrow artifact condition | 0.7487 | 0.1847 |
| F2 | F5 | Waveform-only ECG vs contrast-enhanced condition | 0.8220 | 0.1549 |
| F2 | F6 | Waveform-only ECG vs Gaussian-blurred condition | 0.8063 | 0.1427 |

***Table 4**. Prediction consistency and confidence divergence relative to waveform-only ECG reference F2* ***(Table note:*** *PCS denotes Prediction Consistency Score, where higher values indicate greater agreement in predicted class labels between two feature sets. JSD denotes Jensen–Shannon divergence, where higher values indicate greater divergence in predicted class-probability distributions. The analysis was performed on test samples across all six feature-set prediction files.)*

**Integrated Gradients Results:** Integrated Gradients analysis was performed to examine whether the CNN models attributed their predictions to clinically relevant ECG structures or to artifact-induced visual patterns. The paired comparisons in **Fig. 6**, show how attribution patterns change when non-physiological visual modifications are introduced. Brighter regions in the attribution maps indicate image areas with higher contribution to the model prediction. Average Integrated Gradients maps generated on 20 test samples showed clear differences across three ECG Feature Sets (FS1, FS2 and FS3) when compared to outputs of raw ECG images. In the raw ECG and waveform-only conditions (**Fig. 6**), attribution was broadly distributed across the image, indicating that the CNN used spatial information from multiple ECG regions. The metadata-only condition showed weaker and more localized attribution patterns with higher activation mainly concentrated in limited regions of the image where the patients ID and the ECG specific markers are present. This indicates that even without the actual waveforms present in the images, model still used residual non-waveform information, but the attribution pattern became more restricted and concentrated over the corners when compared with waveform-containing representations. Overall, the IG heatmap results support that the model attribution changes substantially when the ECG waveform is removed leads to the presence of shortcut learning and effect of Clever Hans.

Comparison for class-specific induced feature sets (Feature Set 4,5 and 6) are presented in **Fig. 7**. The red-arrow condition (Feature Set 4) showed reduced and more restricted attribution compared with the raw ECG image, indicating that the attribution pattern changed after insertion of the artificial marker and affected the learning patterns but retained similar performance. The contrast-enhanced condition (Feature Set 5 (**Fig. 7**)) showed a relatively similar but less intense attribution pattern compared with the raw ECG image. It suggests that contrast enhancement altered the strength of attribution while preserving broad spatial activation throughout the whole image which was also witnessed as significant performance increase. The Gaussian-blur condition (Feature Set 6 (**Fig. 7**)) showed a marked reduction in attribution intensity and concentration on waveform regions indicating that blurring substantially changed the attribution distribution and weakened the visible contribution regions which was further evidenced in performance decline.

If the attribution distribution shifts toward the modified or visually altered regions, this suggests that the model may be using artifact-induced cues rather than relying exclusively on ECG waveform morphology. Therefore, these attribution maps provide visual evidence for assessing shortcut learning under controlled artifact-induced conditions.

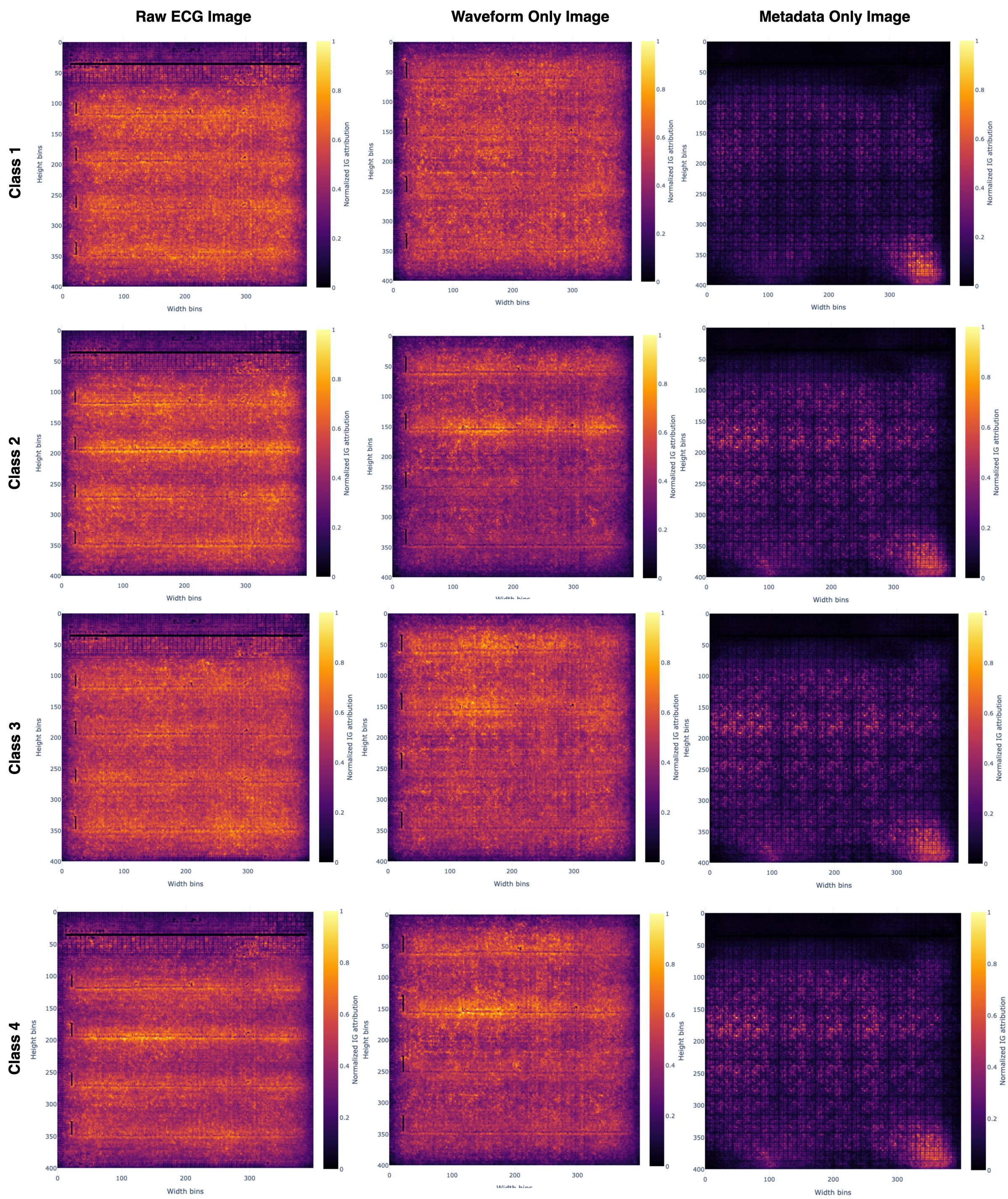

***Fig. 6:*** *Average Integrated Gradients Attribution Maps Across Raw, Waveform-Only, and Metadata-Only ECG Representations*

## Raw vs Red Arrow (Class 1)

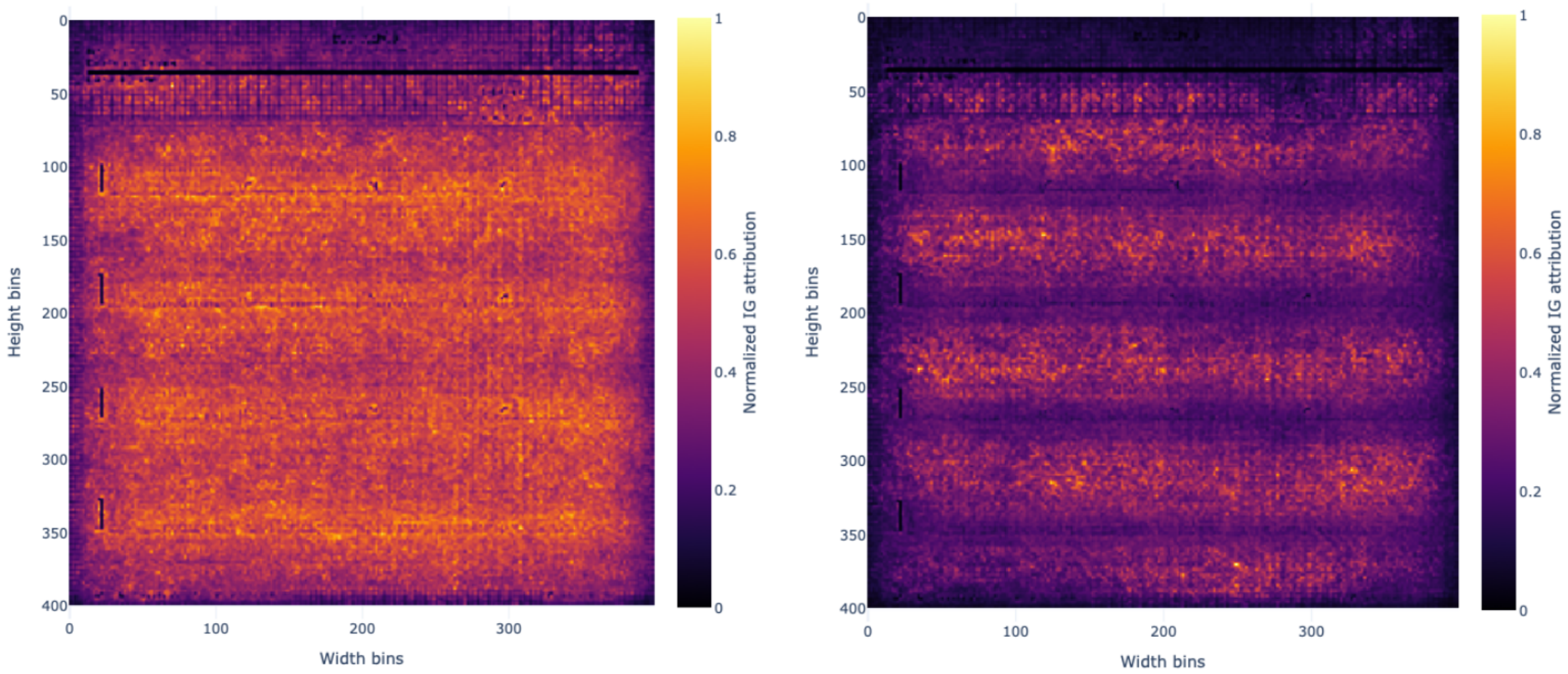

## Raw vs Contrast (Class 2)

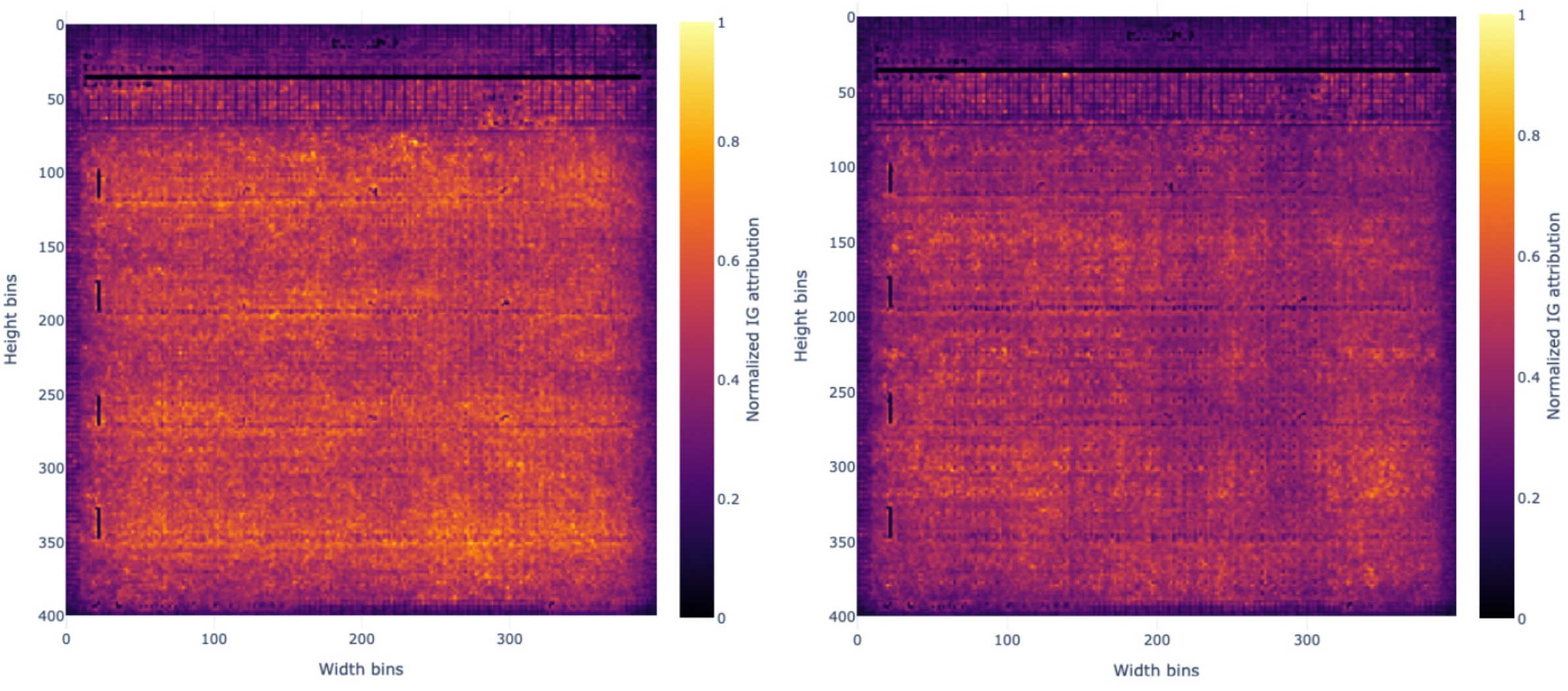

## Raw vs Gaussian Blur (Class 4)

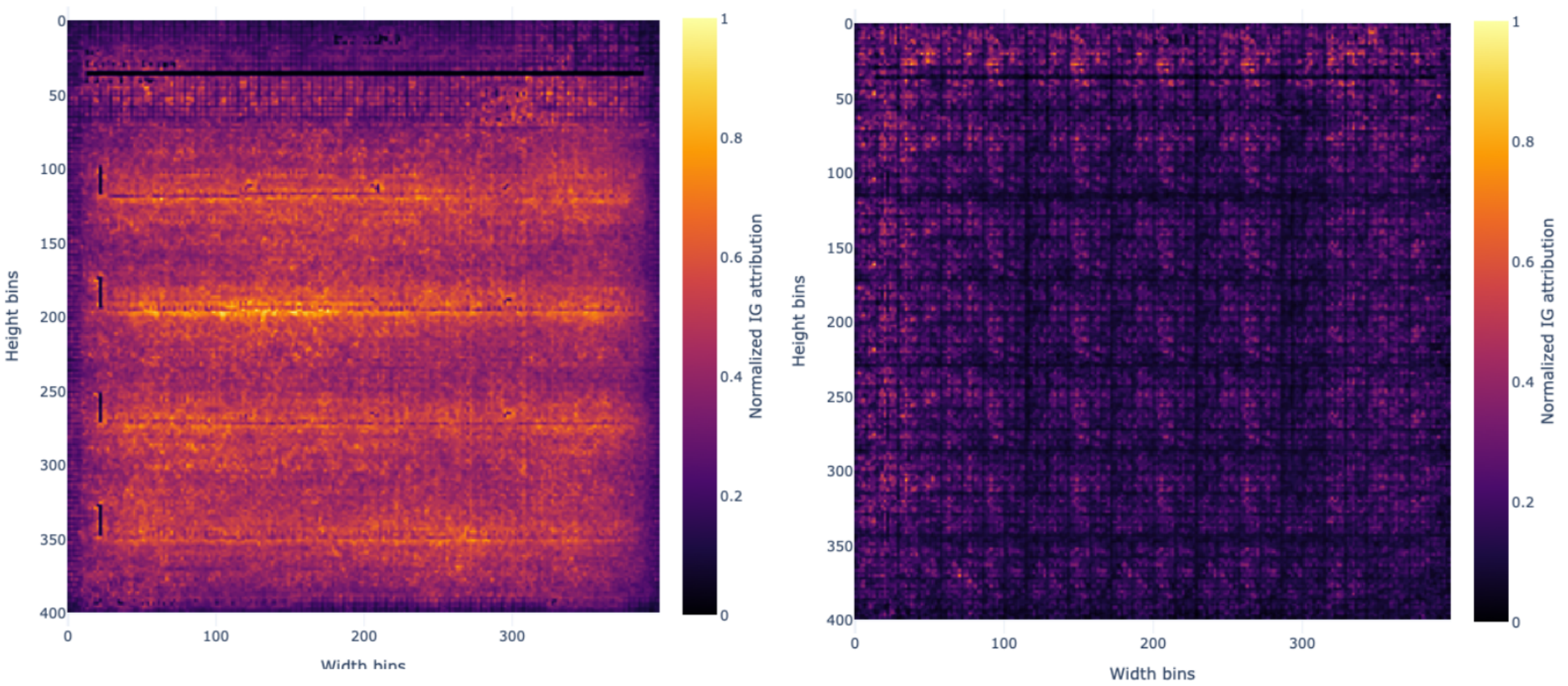

***Fig.** 7: Average Integrated Gradients attribution comparison under artifact-induced ECG conditions.*

**Occlusion Sensitivity Test:** Patch-wise perturbation occlusion sensitivity analysis was performed to quantify the effect of systematic local occlusion of ECG image regions on CNN prediction confidence across different feature-set representations. Considering the experimental setting, the OS was first computed (with the experimental setup provided earlier in the manuscript) separately for each of the four ECG classes within each feature set. The class-wise values were then summarized as mean ± SD across the four classes, to give a compact feature-set-level description of perturbation sensitivity. This aggregation enabled us to compare occlusion behavior across F1–F6 but still preserve the variability seen between diagnostic classes.
The OS results showed in **Tabel 5**, that Feature Set 2 had the highest Mean OS (0.0174 ± 0.0172) and Strong OS (0.0402 ± 0.0394), followed by Feature Set 1 with Mean OS 0.0140 ± 0.0094 and Strong OS 0.0320 ± 0.0221. This suggests that the model confidence was more strongly affected by the local patch removal in the waveform-containing representations, particularly in the waveform-only condition. This fits with the idea that the CNN employed ECG waveform regions when this information was present.
Feature Set 3 had the lowest f1-score for Mean OS (0.0010 ± 0.0021) and Strong OS (0.0016 ± 0.0032), almost no flipping of prediction (0.0008 ± 0.0017). This corroborates the very weak global occlusion response for the metadata-only/waveform-masked representation. But Feature Set 3 also had the highest Top 10% concentration (0.6325 +- 0.4306) and increased localization (0.1777 +- 0.1791), meaning that the small sensitivity was concentrated in limited image regions. This pattern suggests the possible use of residual non-waveform cues as a shortcut.
The artifact-introduced features Feature Set 4, Feature Set 5 and Feature Set 6 had low Mean OS values and low flip rates. Feature Set 4 showed Mean OS 0.0077 ± 0.0084 and flip rate 0.0028 ± 0.0032, showing low sensitivity to perturbation under the red-arrow condition. Feature Set 5 had very low Mean OS (0.0025 ± 0.0041) but high Top 10% concentration (0.5322 ± 0.2156), indicating a highly concentrated but low-magnitude occlusion response for the contrast-enhanced condition. Feature Set 6 demonstrated low Mean OS (0.0093 ± 0.0125) and low flip rate (0.0014 ± 0.0027), indicating limited decision instability under Gaussian blur.
In general, representations that contained waveforms showed stronger OS, while representations that contained only metadata and those induced by artefacts showed weaker but spatially more concentrated responses. This is in line with the potential of shortcut learning, as some feature sets could be able to make stable predictions even with less or altered waveform information. The high concentration values in Feature Set 3 and Feature Set 5 are particularly pertinent to Clever Hans-type behavior, where the model may exploit limited non-physiological cues, rather than relying solely on ECG morphology.

| Feature set | Mean OS | Strong OS | Top 10% concentration | Localization | Flip rate |
|---|---|---|---|---|---|
| F1 | 0.0140 ± 0.0094 | 0.0320 ± 0.0221 | 0.2657 ± 0.0586 | 0.0254 ± 0.0103 | 0.0295 ± 0.0289 |
| F2 | 0.0174 ± 0.0172 | 0.0402 ± 0.0394 | 0.2977 ± 0.1055 | 0.0370 ± 0.0233 | 0.0187 ± 0.0207 |
| F3 | 0.0010 ± 0.0021 | 0.0016 ± 0.0032 | 0.6325 ± 0.4306 | 0.1777 ± 0.1791 | 0.0008 ± 0.0017 |
| F4 | 0.0077 ± 0.0084 | 0.0179 ± 0.0201 | 0.3004 ± 0.0765 | 0.0334 ± 0.0203 | 0.0028 ± 0.0032 |
| F5 | 0.0025 ± 0.0041 | 0.0103 ± 0.0168 | 0.5322 ± 0.2156 | 0.0857 ± 0.0570 | 0.0001 ± 0.0003 |
| F6 | 0.0093 ± 0.0125 | 0.0208 ± 0.0271 | 0.1725 ± 0.1169 | 0.2651 ± 0.4900 | 0.0014 ± 0.0027 |

***Table 5.*** *Compact Occlusion Sensitivity Summary Across Six ECG Feature Sets **(**Table note. Values are reported as mean ± SD across the four ECG classes. Mean OS and Strong OS quantify the magnitude of confidence change under patch occlusion, whereas Top 10% concentration and Localization describe the spatial distribution of sensitivity. High concentration or localization should not be interpreted as strong occlusion dependence unless Mean OS or Strong OS is also sufficiently large.**)***

Along with the factual quantification presented in **Table 5**, visual results of occlusion sensitivity test are presented in **Fig. 8** (Feature Set 1, 2 and 3) and **Fig. 9** (Feature Set 4, 5 and 6). Visual OS heatmaps were further examined to assess how the spatial perturbation response varied across feature-set representations. In Feature Set 1 and Feature Set 2 presented in Fig. 8, sensitivity patterns were distributed across multiple spatial regions, indicating that local occlusion affected prediction confidence in several parts of the image representation. In Feature Set 3, the sensitivity maps were generally weaker and more spatially restricted towards presence of metadata and ECG specific markers. It was consistent with the reduced visual information available after waveform masking. Overall, the comparison shows that the occlusion response differs across image representations, with waveform-containing feature sets producing broader perturbation-sensitive regions than the metadata-only condition.

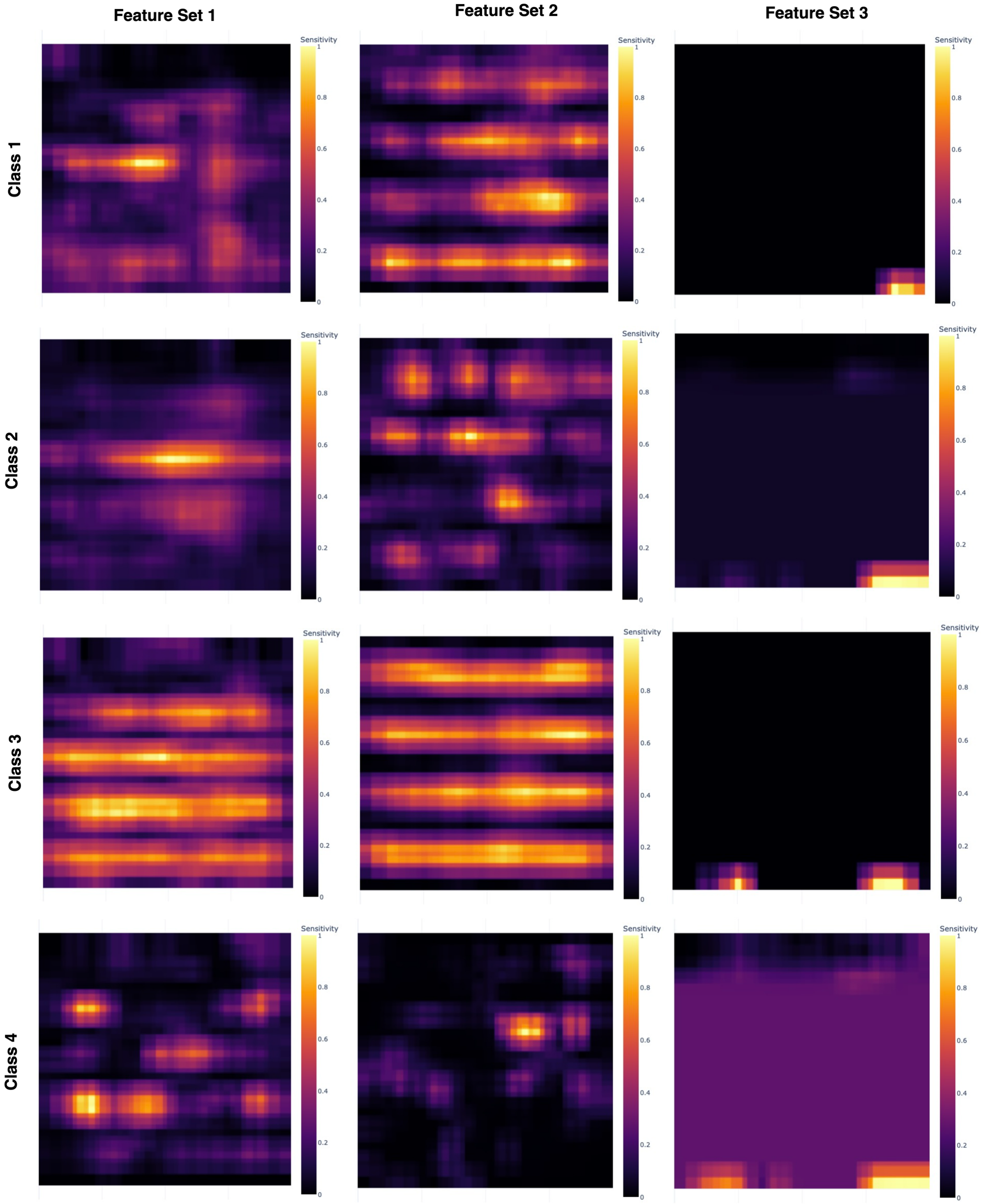

***Fig. 8**: Class-Wise Occlusion Sensitivity Heatmaps Across Raw, Waveform-Only, and Metadata-Only ECG Feature Sets.*

Occlusion sensitivity analysis was used to evaluate whether artifact-induced image modifications altered the perturbation response of the CNN models. For each selected class, the raw ECG condition was compared with its corresponding artifact-induced representation. The red-arrow condition (Feature Set 4, altered class 1) assessed sensitivity to an explicit artificial marker, the contrast-enhanced condition (Feature Set 5, Altered Class 3) assessed intensity-related perturbation response, and the Gaussian-blurred condition (Feature Set 5, Altered Class 4) assessed image-quality-related perturbation response. Differences between the raw and artifact-induced heatmaps indicate that the spatial distribution of occlusion sensitivity changes when non-physiological visual modifications are introduced. The heatmap provided perturbation-based evidence that model confidence was influenced by artifact-induced visual conditions rather than being driven only by ECG morphology.

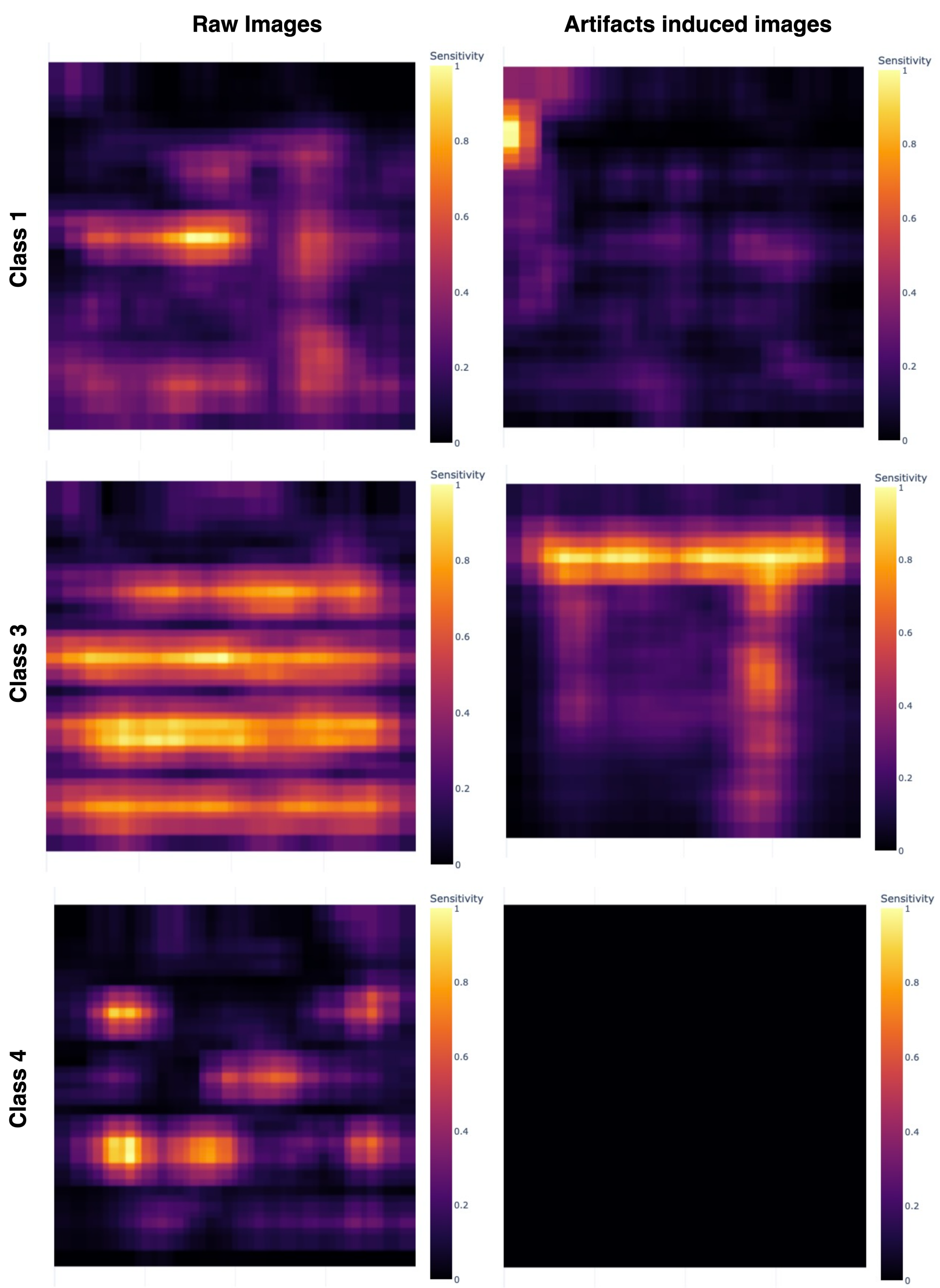

*Fig. 9. Class-specific occlusion sensitivity comparison between raw ECG images and artifact-induced ECG representations.*

**Discussion in Relation to Previous Studies**

The representation-dependent behavior observed across the six feature sets is consistent with a growing body of evidence indicating that high predictive performance in medical image classification does not, by itself, establish that a model has learned clinically meaningful morphology. Geirhos et al. [9] described this behaviour as shortcut learning, in which decision rules that perform well on standard benchmarks fail to transfer to more demanding testing conditions, whereas Lapuschkin et al. [13] demonstrated through relevance heatmaps that a classifier may attain strong benchmark accuracy by attending to a source tag rather than to the object of interest. Comparable dependence on non-target information has been reported in general computer vision, where classifiers trained exclusively on background regions retained substantial recognition accuracy [14], and where models trained on familiar acquisition contexts generalized poorly to unseen environments [15]. The present findings extend these observations to ECG report images, in which waveform morphology and non-physiological report context coexist within a single input and are ordinarily inseparable during conventional training.

The retention of classification performance under the waveform-masked condition parallels the observation of Maguolo and Nanni [21], who reported that COVID-19 classifiers achieved comparable accuracy when the lung fields were replaced by zero-intensity regions, and the findings of DeGrave et al. [20], whose saliency evidence indicated reliance on laterality markers and other acquisition-related cues rather than on pulmonary pathology. Arias-Londoño and Godino-Llorente [17] reached an equivalent conclusion using Bayesian deep learning, attributing apparent diagnostic competence to dataset-specific confounders. Analogous dependencies have been documented outside imaging: Perlich et al. [18] identified predictive information carried by patient identifiers in the KDD Cup breast cancer challenge, and Raimondi et al. [19] showed that cancer driver-variant predictors principally recognized driver genes rather than variant-level functional effects, with performance decreasing markedly once the dataset construction bias was mitigated. At the level of study design, Roberts et al. [7] reported that a substantial proportion of published COVID-19 models were affected by methodological or dataset biases that limited their clinical applicability. Collectively, these studies indicate that shortcut reliance is neither modality-specific nor architecture-specific but arises whenever a dataset offers a discriminative signal that is easier to learn than the intended one.

Within the ECG image literature, several studies have reported very high classification accuracy on the publicly available corpus employed in the present work [26, 27]. Sadad et al. [22] reported an accuracy of 98.39% using a lightweight attention-based convolutional network, and Alsayat et al. [24] reported an F1 score of 0.9651 for an ensemble of Inception, MobileNet and NASNetLarge. Li et al. [28] and Nawaz et al. [29] reported comparable performance for image-based ECG interpretation using deep convolutional architectures, while Ao and He [23] applied Grad-CAM and observed that attribution maps highlighted regions conventionally used by human interpreters. These studies evaluated performance using conventional metrics computed on data drawn from a single acquisition setting, and none of them examined whether the reported accuracy would persist once waveform information was removed or once class-specific artefacts were introduced. Czerwinski et al. [25] provided partial evidence in this direction, reporting a pronounced decrease in performance when models trained on one ECG scan dataset were evaluated on another, which is consistent with dependence on dataset-specific rather than physiological characteristics. The controlled feature-set design adopted in the present study addresses this gap directly, since Feature Set 2 and Feature Set 3 separate morphology from report context, and Feature Set 4 to Feature Set 6 introduce class-specific artefacts of known origin, thereby allowing the source of predictive information to be attributed rather than inferred.

The attribution and perturbation evidence reported above should nevertheless be interpreted with the caution advocated in the explainability literature. Reyes et al. [6] noted that interpretability outputs support, but do not by themselves establish, the clinical validity of a model, and Anders et al. [16] demonstrated that Clever Hans strategies can be identified and partially removed through explanation-guided model correction. Bassi et al. [30] further showed that optimizing layer-wise relevance propagation attributions improves robustness to background bias, indicating that explanation methods may serve a corrective as well as a diagnostic function. Considered together with the results reported here, this literature suggests that representation-level testing and explanation-based auditing are complementary rather than interchangeable procedures, and that both are required before an ECG image classifier trained on report images can reasonably be regarded as clinically dependable.

## Conceptual Framework for Assessing Shortcut Learning and Clever Hans Effect

The conceptual framework proposed in Fig.10, characterizes shortcut learning as a model behavior problem that is representation dependent. Raw input images may contain relevant features and irrelevant cues, such as contextual, confounding, or shortcut information. The model processes such mixed representations and produces measurable behavioral outputs such as performance variation, prediction consistency and confidence divergence across representations. These outputs are then backed by XAI evidence, where visual and factual analyses help to determine whether the decision process is guided by task-relevant features or by shortcut cues. The XAI evidence feedback informs the final interpretation of model behavior, separating relevant-feature driven learning from shortcut driven behavior. Within this framework, Clever Hans-type behavior is seen as a case in which the model seems to be performing well but is in fact relying on irrelevant or non-causal cues rather than meaningful features.

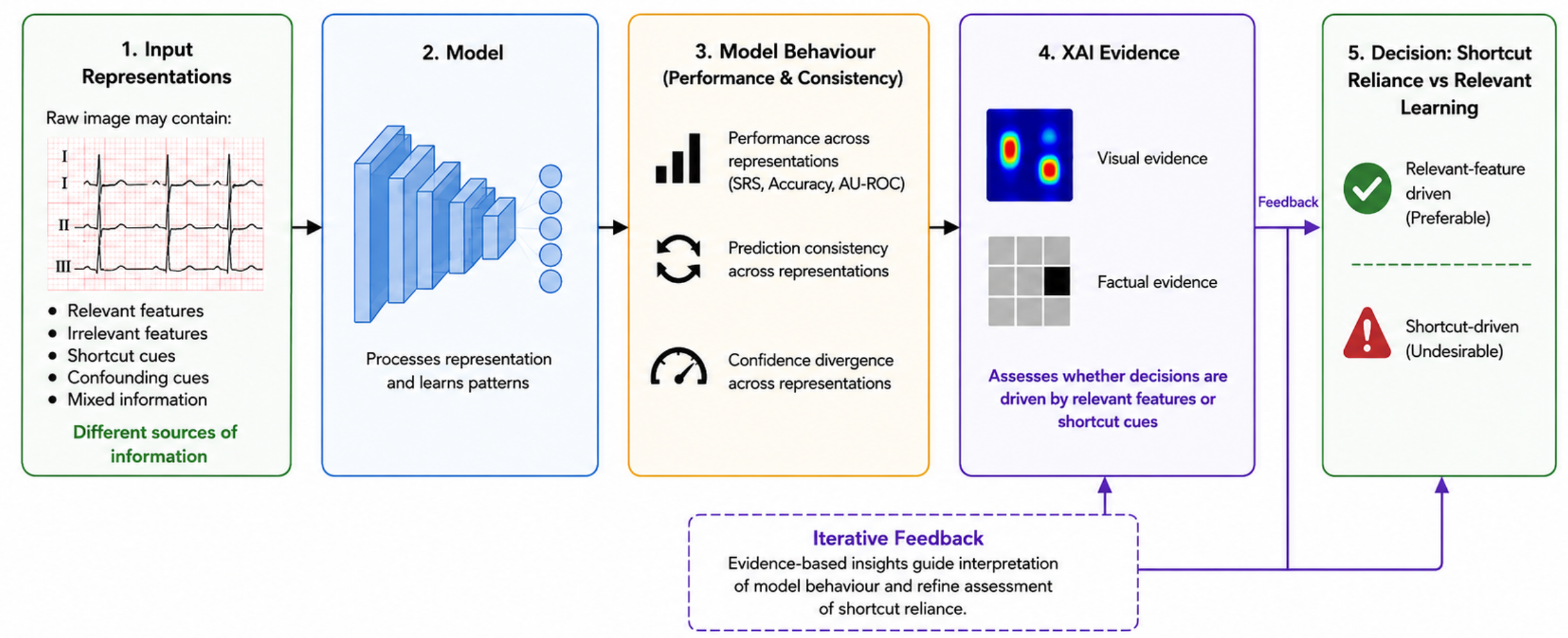

***Fig. 10**: Conceptual Framework for Assessing Shortcut Learning and Clever Hans Effect in AI Models.*

## Implications for Practitioners

Practitioners should interpret high ECG-AI accuracy with caution when models are trained on image-based ECG reports. Model validation should consider whether predictions are driven by clinically relevant waveform morphology or by non-physiological visual cues. XAI outputs should be used as supporting evidence, not as proof of clinical reasoning. Before clinical deployment, AI-based clinical decision support systems should be tested under representation changes, artifact variation, and external validation settings.

## Limitations of this study

This study has few limitations. First, the analysis was performed on a single publicly available ECG image dataset; therefore, external validation on independent datasets is required. Second, the experiments used a CNN-based classifier, and the observed shortcut behavior may differ with other model architectures. Third, the artifact-induced feature sets were designed for controlled shortcut assessment and may not capture all real-world ECG reporting or acquisition artifacts. Fourth, XAI and occlusion-based analyses provide useful evidence of model behavior, but they do not prove causal clinical reasoning. The findings should be interpreted as evidence of potential shortcut reliance rather than definitive proof of model failure.

## Conclusion and Future Directions

To conclude, this study provides substantial evidence of shortcut learning and CH effect in a commonly used publicly available dataset. This study demonstrates why it is important to have correct interpretation of models, while deploying in real-time scenarios. External assessment should be performed on independent datasets to observe whether the observed model behavior generalizable beyond this current dataset. Explainable AI techniques and transparent learning should be common practice when dealing with sensitive applications. Artifacts-robust training and validation strategies should be explored including pre-processing techniques, artifacts removal, presentation level testing and prospective evaluation with domain knowledge.

### Author Contributions

**Research conceptualization and design:** Abhay Kumar Pathak, Mrityunjay Chaubey, Manjari Gupta, Deepti Mishra; **Supervision and Validation**: Manjari Gupta, Deepti Mishra; **Project Administration:** Manjari Gupta, Deepti Mishra; **Collection and/or assembly of data**: Abhay Kumar Pathak; **Data analysis and interpretation:** Abhay Kumar Pathak, Mrityunjay Chaubey; **Visualization:** Abhay Kumar Pathak, Mrityunjay Chaubey; **Writing- Original Draft:** Abhay Kumar Pathak, Mrityunjay Chaubey; **Resources:** Manjari Gupta, Deepti Mishra; **Writing- Review & Editing:** Mrityunjay Chaubey, Manjari Gupta, Deepti Mishra.

### Funding

NA

### Conflict of interest

The authors declare that there are no conflicts of interest related to this work.

### Data Availability Statement

The dataset supporting the findings of this study is the publicly available ECG Images Dataset of Cardiac Patients, contributed by Ali Haider Khan and Muzammil Hussain, and accessible through Mendeley Data. The dataset consists of 12-lead ECG report images created under the auspices of the Ch. Pervaiz Elahi Institute of Cardiology,

Multan, Pakistan, and is intended to support cardiovascular disease research. The ECG images are organized according to expert-defined diagnostic categories, and the binary labels used in this study were assigned from the dataset's class-wise annotations according to the study-specific label-mapping protocol. The dataset is available under a CC BY 4.0 licenseand can be accessed via DOI: 10.17632/gwbz3fsgp8.2.